\documentclass[journal]{IEEEtran}
\usepackage[T1]{fontenc}
\usepackage[latin9]{inputenc}
\usepackage{amsmath}
\usepackage{amssymb}
\usepackage{amsthm,bm}
\usepackage{url}
\usepackage{cite}
\usepackage{babel,verbatim,balance}
\usepackage[rgb]{xcolor}
\usepackage{graphics, graphicx,subcaption}
\usepackage{acronym}
\usepackage[bookmarks,colorlinks]{hyperref}
\usepackage{setspace,gensymb}

\acrodef{adc}[ADC]{Analog-to-Digital Convertor}
\acrodef{dac}[DAC]{digital-to-analog convertor}
\acrodef{cs}[CS]{Compressed Sensing}
\acrodef{em}[EM]{ElectroMagnetic}
\acrodef{dtft}[DTFT]{discrete-time Fourier transform}
\acrodef{dnn}[DNN]{deep neural network} 
\acrodef{csi}[CSI]{Channel State Information}
\acrodef{map}[MAP]{maximum a-posteriori probability}
\acrodef{snr}[SNR]{Signal-to-Noise Ratio}
\acrodef{sinr}[SINR]{signal-to-interference-and-noise ratio}
\acrodef{bs}[BS]{Base Station} 
\acrodef{iot}[IOT]{Internet of Things}
\acrodef{mimo}[MIMO]{Multiple-Input Multiple-Output}
\acrodef{mse}[MSE]{Mean-Squared Error}
\acrodef{pdf}[PDF]{probability density function}
\acrodef{rv}[RV]{random variable}
\acrodef{tdd}[TDD]{time division duplexing}
\acrodef{rs}[RS]{Reed-Solomon}
\acrodef{lti}[LTI]{linear time-invariant}
\acrodef{wss}[WSS]{wide-sense stationary}
\acrodef{psd}[PSD]{power spectral density}
\acrodef{ser}[SER]{symbol error rate} 
\acrodef{ber}[BER]{bit error rate} 
\acrodef{isi}[ISI]{intersymbol interference}  
\acrodef{awgn}[AWGN]{additive white Gaussian noise} 
\acrodef{ut}[UT]{User Terminal} 
\acrodef{dc}[DC]{Direct Current} 
\acrodef{aoa}[AoA]{Angle of Arrival} 
\acrodef{mmw}[mmWave]{millimeter wave}
\acrodef{ris}[RIS]{Reconfigurable Intelligent Surface} 
\acrodef{hris}[HRIS]{Hybrid Reconfigurable Intelligent Surface} 
\acrodef{dma}[DMA]{Dynamic Metasurface Antenna}

\begin{document}
\title{
Hybrid Reconfigurable Intelligent Metasurfaces: Enabling Simultaneous Tunable Reflections and Sensing for 6G Wireless Communications
}
\author{George C. Alexandropoulos,~\IEEEmembership{Senior Member,~IEEE}, Nir Shlezinger,~\IEEEmembership{Member,~IEEE},\\ Idban Alamzadeh,~\IEEEmembership{Graduate~Student~Member,~IEEE,} Mohammadreza F. Imani,~\IEEEmembership{Member,~IEEE},\\ Haiyang Zhang,~\IEEEmembership{Member,~IEEE}, and Yonina C. Eldar,~\IEEEmembership{Fellow,~IEEE}
\thanks{
The work has been supported by the EU H2020 RISE-6G project under Grant Agreement 101017011, and the SNS JU project TERRAMETA under the EU's Horizon Europe research and innovation programme under Grant Agreement No 101097101, including top-up funding by UKRI under the UK government's Horizon Europe funding guarantee.}
}

\maketitle
\begin{abstract}
The latest discussions on the upcoming sixth Generation (6G) of wireless communications are envisioning future networks as a unified communications, sensing, and computing platform. The recently conceived concept of the smart radio environment, enabled by Reconfigurable Intelligent Surfaces (RISs), contributes towards this vision offering programmable propagation of information-bearing signals. Typical RIS implementations include metasurfaces with almost passive unit elements capable of reflecting their incident waves in controllable ways. However, this solely reflective operation induces significant challenges for the RIS optimization from the wireless network orchestrator. For example, RISs lack information to locally tune their reflection pattern, which can only be acquired by other network entities, and then shared with the RIS controller. Furthermore, channel estimation, which is essential for coherent RIS-empowered communications, is challenging with the available RIS designs. This article reviews the emerging concept of Hybrid reflecting and sensing RISs (HRISs), which enables metasurfaces to reflect the impinging signal in a controllable manner, while simultaneously sensing a portion of it. The sensing capability of HRISs facilitates various network management functionalities, including channel parameter estimation and localization, {while giving rise to potentially computationally autonomous and self-configuring metasurfaces}. We discuss a hardware design for HRISs and detail a full-wave electromagnetic proof of concept. The distinctive properties of HRISs, in comparison to their solely reflective counterparts, are highlighted and a simulation study evaluating their capability for performing full and parametric channel estimation is presented. Future research challenges and opportunities arising from the HRIS concept are also included.
\end{abstract}

\section{Introduction}
The potential of \acp{ris} for programmable \ac{em} wave propagation has recently motivated extensive academic and industrial interests as a candidate smart connectivity paradigm for the sixth Generation (6G) of wireless communications \cite{Samsung,WavePropTCCN}. 
The \ac{ris} technology~\cite{huang2019reconfigurable_all}, which typically refers to artificial planar structures with almost passive electronic circuitry, is envisioned to be jointly optimized with conventional transceivers~\cite{wu2021intelligent_ALL} in order to significantly boost wireless communications in terms of coverage, spectral and energy efficiency, reliability, and security, while satisfying regulated EM field emissions.



The typical unit element of an RIS is the meta-atom, which is usually fabricated to realize multiple programmable states corresponding to distinct EM responses. By externally controlling these states, various reflection and scattering profiles can be emulated \cite{Tsinghua_RIS_Tutorial_ALL}. The RIS implementations up to date do not include any power amplification circuitry, hence, comprise of metasurfaces that can only act as tunable reflectors.
%
While passive \acp{ris} enable programmable wireless propagation environments, their purely reflective operation induces notable challenges when considered for wireless networking. For instance, 
{the fact that \acp{ris} cannot sense the environment implies that they lack information to dynamically configure their reflection pattern. For this reason, they are typically externally configured by a dedicated network entity, or a \ac{bs}, via dedicated control links, which in turn complicates their deployment and network-wise management \cite{RISE6G_COMMAG_ALL,RIS_challenges}. Furthermore,} 
the inclusion of an \ac{ris} implies that a signal transmitted from a \ac{ut} to a \ac{bs} undergoes at least two channels: the \ac{ut}-\ac{ris} and \ac{ris}-\ac{bs} channels. Estimating these individual channels is a challenging task due to the reflective nature of \acp{ris} \cite{Tsinghua_RIS_Tutorial_ALL}, which significantly limits the ability to reliably communicate in a coherent manner. 
In addition, solely reflective \acp{ris} impose challenges on wireless localization \cite{Keykhosravi2022infeasible_all}. To overcome them, it was recently proposed to equip \acp{ris} with a dedicated external device comprising a few reception  Radio Frequency (RF) chains \cite{taha2021enabling_ALL}, enabling low-cost signal reception and relevant processing capabilities. 

Radiating metasurfaces have recently emerged as a promising technology for realizing low-cost and low-power extremely large \ac{mimo} antenna arrays~\cite{shlezinger2020dynamic_all}. 
\acp{dma} pack large numbers of controllable radiative meta-atoms that are coupled to one or several waveguides, resulting in \ac{mimo} transceivers with advanced analog processing capabilities. 
While the implementation of \acp{dma} differs from reflective \acp{ris}, the similarity in their unit element structures indicates the feasibility of designing hybrid reflecting and sensing meta-atoms. Metasurfaces consisting of such meta-atoms can reflect their impinging signal, while simultaneously measuring a portion of it \cite{alamzadeh2021reconfigurable_all}. Such Hybrid reflecting and sensing RISs (HRISs) bear the potential of significantly facilitating RIS orchestration, without notably affecting 
the coverage extension advantages offered by solely reflective RISs. 

In this article, we review the emerging concept of HRISs and discuss their possible prominent applications for future wireless communications. We first review the possible configurations for hybrid meta-atoms and showcase the feasibility of HRISs, by describing an implementation of simultaneously reflecting and sensing meta-atoms based on \cite{alamzadeh2021reconfigurable_all}, accompanied with full-wave EM simulations. We proceed with the investigation of the potential of HRISs for wireless communications, demonstrating their ability to achieve desired reflection patterns, while using their sensing capability to locally estimate angles-of-arrival (AoAs), enabling their sensing autonomicity and consequently self-configuration.
Then, we highlight their distinctive properties in contrast to solely reflective \acp{ris} and active relays. In particular, we show that, as opposed to active relays, HRISs preserve the power-efficient reflective operation of conventional RISs, as they do not require any transmission circuitry, and can perform sensing with a very small number of reception RF chains. Compared to equipping conventional RISs with dedicated receivers~\cite{taha2021enabling_ALL}, HRISs allow to sense the signals observed at each reflective element, and implement reception directly at the metasurface. A simple model for the dual HRIS operation is introduced, facilitating a representative simulation study of the HRIS capability for signal parameter sensing and channel estimation. We conclude with a description of some research challenges and opportunities with HRISs, including key experimental directions which are expected to further unveil their potential role in 6G wireless networks. 

\section{HRIS Design and Proof Of Concept}
\label{sec:Hybrid} 
A variety of RIS implementations has recently been introduced~\cite{wu2021intelligent_ALL}, ranging from metasurfaces that manipulate wave propagation in rich scattering environments to improve the received signal strength, to those that realize desired anomalous reflection beyond Snell's law. 
In all fabricated designs, the RISs are incapable of any sort of sensing of the impinging signal. The concept of RISs realizing sensing and communicating was firstly envisioned in \cite{alamzadeh2021reconfigurable_all}, where an implementation with sensing capabilities at the meta-atom level was demonstrated. 
We next present HRISs incorporating this hybrid meta-atom design, and that leverage previous works on DMAs \cite{shlezinger2020dynamic_all} to integrate sensing into each meta-atom, without losing their reconfigurable reflection functionalities.

\begin{figure}
    \centering
    \includegraphics[width=\columnwidth]{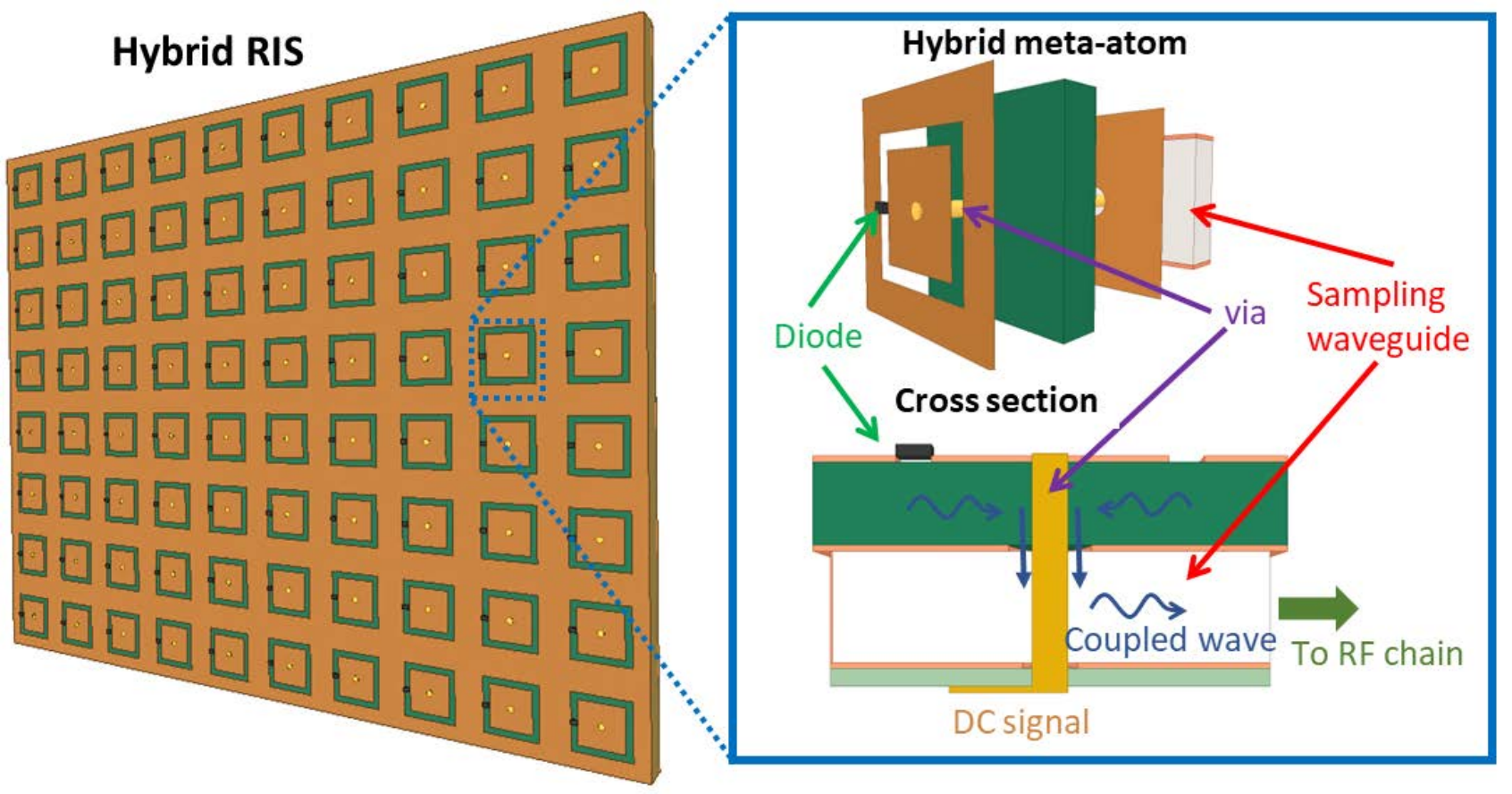}
    \vspace{-0.2cm}
    \caption{\small{Illustration of the HRIS and the constitutive hybrid meta-atom design proposed in \cite{alamzadeh2021reconfigurable_all}. The layers of the meta-atom on the top right are artificially separated for visualization purposes. The cross section of the metasurface and the coupled-wave signal path are depicted on the bottom-right drawing.}}
    \label{fig:HybridAtom_v01}
\end{figure}

\smallskip
{\bf Configurations of Hybrid Meta-Atoms}: From a hardware perspective, one can propose two different configurations that allow for meta-atoms to both reflect and sense. The first consists of hybrid meta-atoms, which simultaneously reflect a portion of the impinging signal, while enabling another portion to be sensed \cite{alamzadeh2021reconfigurable_all}. 
The second configuration uses meta-atoms that reconfigure between near-perfect absorption and reflection. In the latter case, each meta-atom has two modalities \cite{ABSence_all}: it can either sense the received signal (absorption mode), or, once necessary information is extracted, reflect the signal towards a desired direction (reflection). While the configurations vary in their implementation, in terms of modeling, the first configuration may be viewed as a generalization of the second, since a hybrid meta-atom can be tuned to fully absorb or sense. Accordingly, we focus henceforth on the first case relying on hybrid meta-atoms, as illustrated in Fig.~\ref{fig:HybridAtom_v01}. Such HRISs are realized by adding a waveguide to couple to each or groups of meta-atoms. Each waveguide can be connected to a reception RF chain, allowing the HRIS to locally process a portion of the received signals in the digital domain. However, the elements' coupling to waveguides implies that the incident wave is not perfectly reflected. In fact, the ratio of the reflected energy to the sensed one is determined by the coupling level. By keeping this waveguide near cutoff, we can reduce its footprint, while also reducing coupling to the sampling waveguide.


{\bf Design Considerations}: An important consideration in the HRIS design is the inter-element spacing. To accurately detect the phase front of the incident wave, the spacing of adjacent meta-atoms needs to be smaller than half a wavelength. However, this imposes constraints on the meta-atom size, which is accommodated via the multi-layer structure in Fig.~\ref{fig:HybridAtom_v01}. Another important factor to consider is the necessary circuitry to detect the signal coupled from each meta-atom; in particular, each waveguide should be connected to an RF chain. Since the incident wave on the HRIS may couple to all sampling waveguides (with different amplitudes), we can think of the combination of the metasurface and the sampling waveguides as a receiver structure with equivalent analog combining. 
\begin{figure*}
    \centering
    \begin{subfigure}[h]{0.4\textwidth}
        \includegraphics[width=\textwidth]{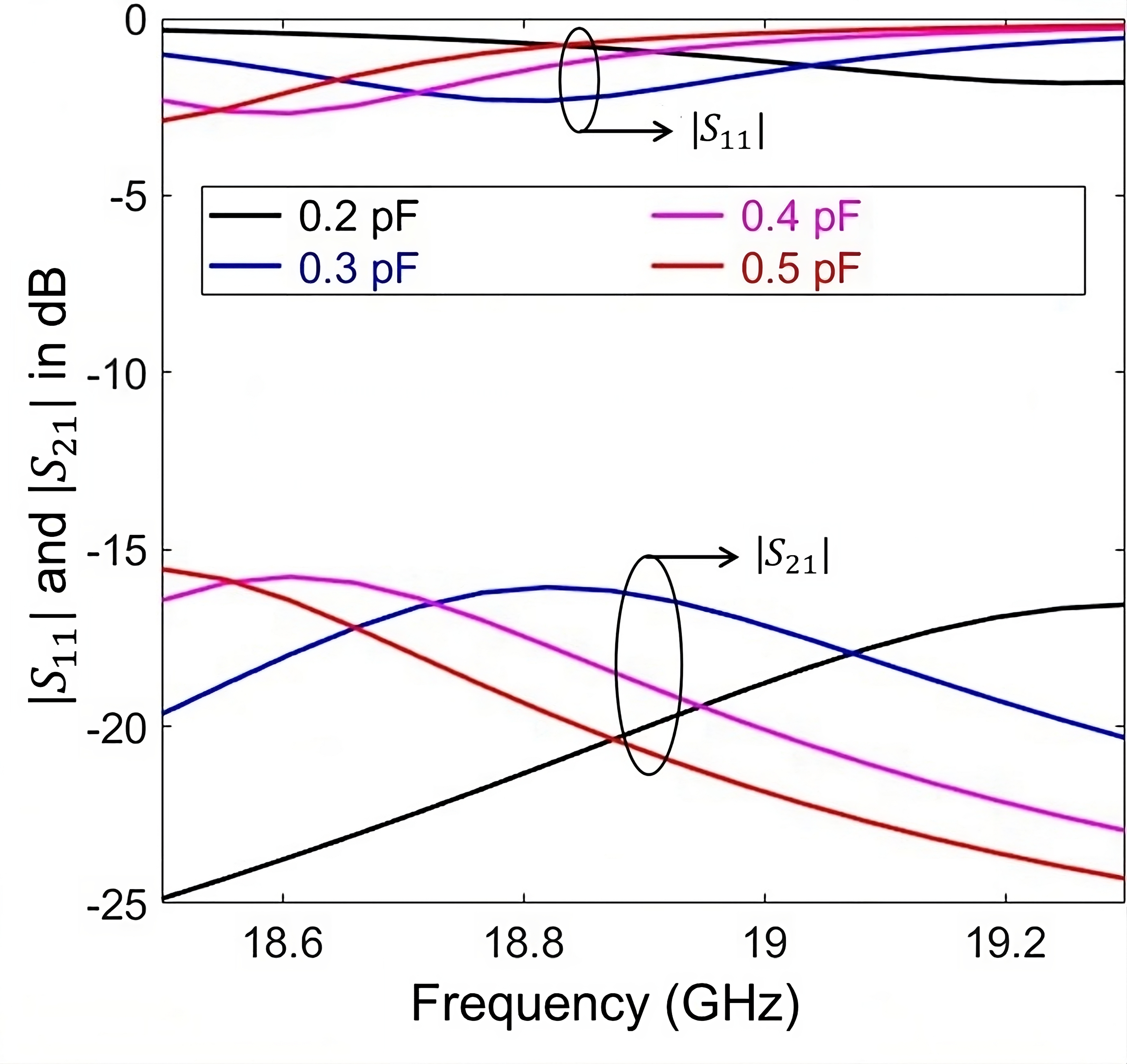}
        \caption{\small{Amplitudes of $S_{11}$ and $S_{21}$.}}
        \label{fig:S_amplitudes}
    \end{subfigure}
    ~ 
			\quad\quad\quad\quad
    \begin{subfigure}[h]{0.4\textwidth}
        \includegraphics[width=\textwidth]{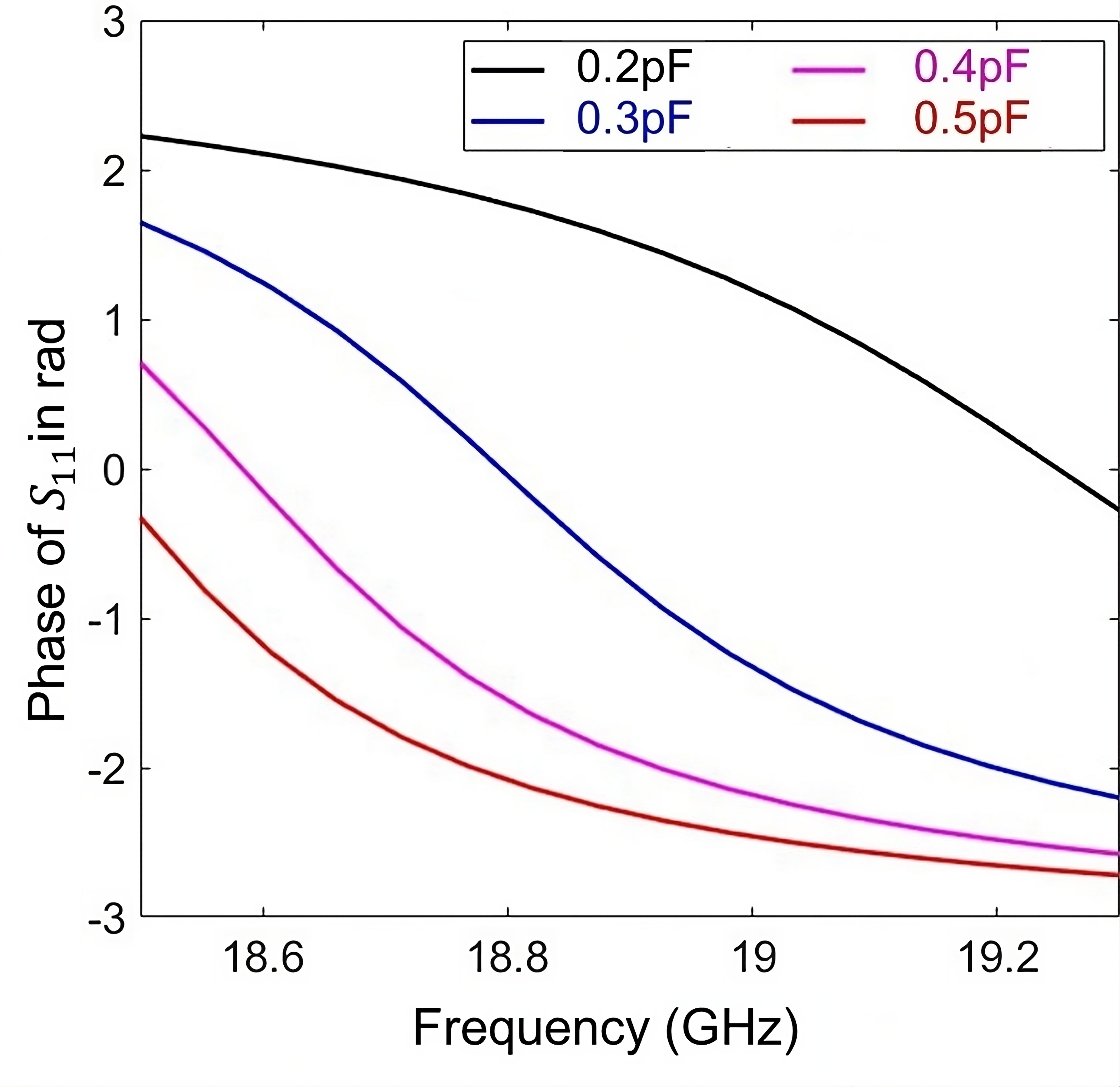}
        \caption{\small{Phase of $S_{11}$.}}
        \label{fig:AN_ONLY_B}
    \end{subfigure}
    \caption{\small{Full-wave EM simulation of a single hybrid meta-atom \cite{alamzadeh2021reconfigurable_all} around the operating frequency $19$ GHz. (a) The $S_{11}$ parameter indicates the reflection from the meta-atom, while the $S_{21}$ parameter represents the coupling to its sampling waveguide. The dip in the magnitude of $S_{11}$ indicates the resonance of this element that changes as a function of the effective capacitance of the deployed varactor diode. The magnitude of the coupling $S_{21}$ to the sensing waveguides has been kept to be below $-15$ dB to ensure that ample power is reflected by this meta-atom. It is depicted that the coupling changes as a function of the effective capacitance of the varactor diode. (b) The corresponding reflection phase of the hybrid meta-atom changes as a function of the effective capacitance of the varactor diode. This change in phase can be used to generate different reflection patterns, enabling the redirection of the impinging signal on the meta-atom towards different directions.}}
\label{fig:AN_ONLY_Both}
\vspace{-0.2cm}
\end{figure*}

The sensing circuitry, however, adds another factor to consider when designing an HRIS: while closer element spacing improves the ability to direct the reflected signal to a desired direction (with smaller sidelobes) and increases the accuracy of detecting incident wave fronts, close element spacing increases the total number of meta-atoms, and consequently, the cost, and potentially, the power consumption. Nonetheless, this can be balanced by limiting the number of RF chains---which are connected to the waveguides and not directly to each meta-atom--- and profiting from each meta-atom's controllable analog processing nature. The latter feature also contributes to reducing the cost and power consumption of the introduced sensing circuitry, which does not appear in conventional \acp{ris} that are incapable of sensing. 

\smallskip
{\bf Implementation}: 
Most RIS elements in available designs are resonators, which can be easily modified to couple to a waveguide. For example, the meta-atom in~\cite{sleasman2016microwave_all}  
requires a via to deliver the \ac{dc} signal to this element, in order to tune the switchable component. To eliminate the undesired coupling from the RF signal to the DC bias line, this via is usually terminated in a radial stub. In the hybrid meta-atom implementation proposed in \cite{alamzadeh2021reconfigurable_all}, we utilized this coupling to implement sensing of the incident signal. One can envision a design where the via is attached to two copper traces, one to sample the signal and another to transfer the DC signal (which is connected to a radial stub), though confining the coupled RF signal to such a copper trace may be challenging.
A Substrate Integrated Waveguide (SIW), which is effectively a rectangular waveguide, is used to capture the sampled wave, as shown in Fig.~ \ref{fig:HybridAtom_v01}. By changing the annular ring around the coupling via or the geometrical size of the SIW, the HRIS can realize different coupling strengths.

\smallskip
{\bf Full-Wave EM Simulations}: The hybrid meta-atom illustrated in Fig.~\ref{fig:HybridAtom_v01} is 
loaded by a varactor, whose effective capacitance is changed by an external \ac{dc} signal, consequently changing the reflection phase of its impinging wave. By properly designing the phase variation along the HRIS, the reflected wave can be steered towards desired directions. 
To evaluate the ability of our HRIS design to simultaneously reflect while sensing its incident signal, we next present a full-wave simulation study using Ansys HFSS. We first evaluate a hybrid meta-atom operating at $19$ GHz, focusing on its reflection and coupled components in Fig.~\ref{fig:AN_ONLY_Both}. It is demonstrated that this element can be tuned to reflect and sense different portions of its impinging signal (Fig.~\ref{fig:S_amplitudes}), while applying controllable phase shifting to the reflected signal (Fig.~\ref{fig:AN_ONLY_B}). Additional investigations of a full HRIS comprised of such hybrid meta-atoms, further showcasing the design's validity, are available in~\cite{alamzadeh2021reconfigurable_all}.

\section{HRISs for Smart Wireless Environments}
\label{sec:Comm}
In this section, we discuss how the concept of HRISs, 
via the aforedescribed implementation guidelines, can be exploited to facilitate wireless operations. We first describe the envisioned HRIS-empowered wireless communications paradigm, and then, present a simple model encapsulating the simultaneous tunable reflection and sensing operations of HRISs. Finally, some numerical results corroborating their potential for AoA recovery and full channel estimation are presented.
\begin{figure*}
    \centering
    \includegraphics[width=\linewidth]{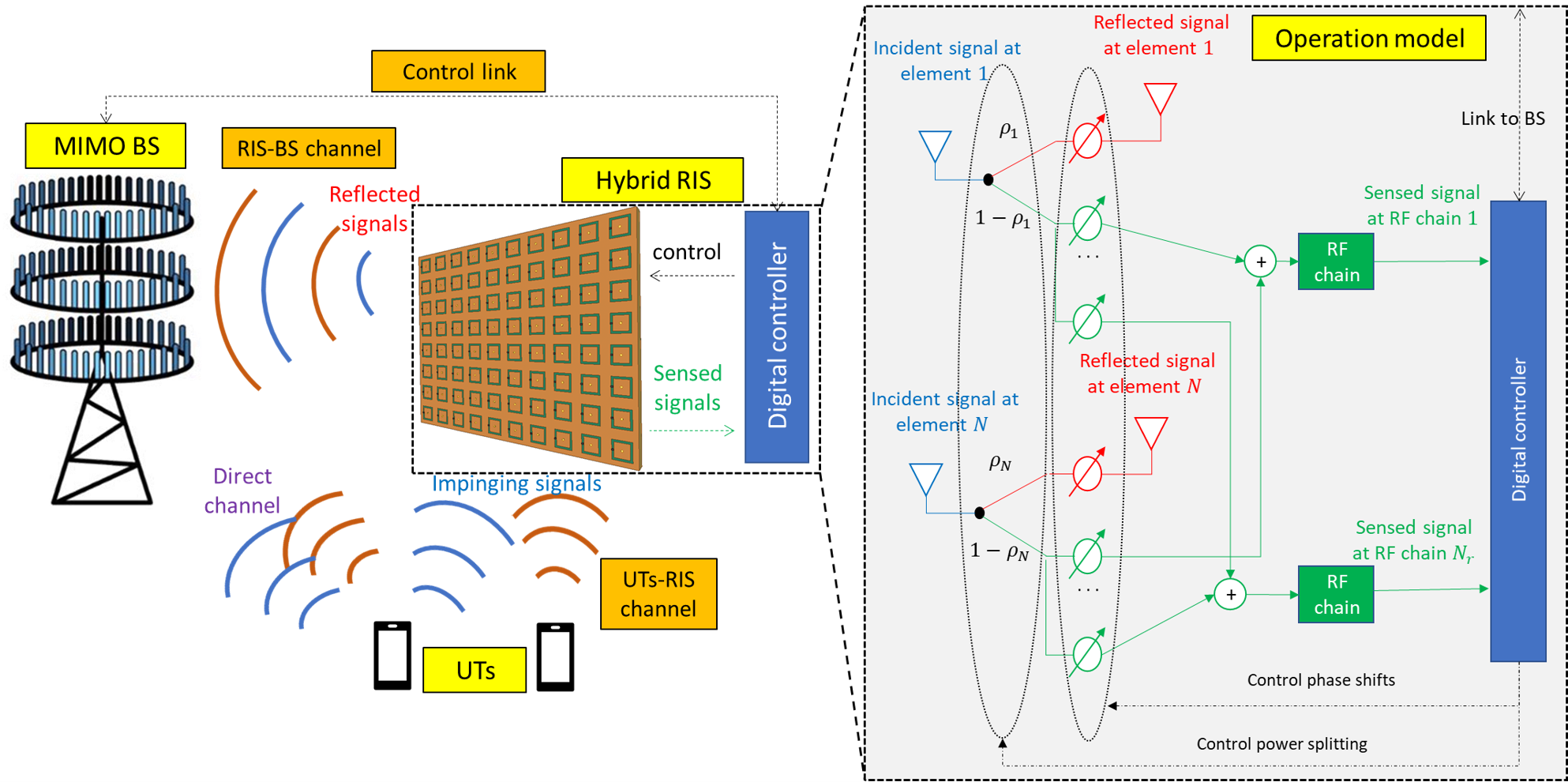}
    \caption{\small{The uplink of multi-user \ac{mimo} systems incorporating the proposed HRIS structure that also senses a portion of the impinging UT signals (left), along with the operation model of the HRIS (right). In this model, the HRIS consists of $N$ hybrid meta-atoms. The incident signal at each meta-atom is split into a portion which is reflected (after tunable phase shifting) in the environment, while the remainder of the signal is sensed and processed locally at the metasurface. Parameter $\rho_n$ ($n=1,2,\ldots,N$) represents the coupling coefficient (i.e., the power splitting ratio between reflection and sensing operations) for each $n$-th meta-atom. The received portion of the impinging signal undergoes each element's controllable sensing response and is collected at the $N_r$ reception RF chains connected to the $N$ sampling waveguides ($N\gg N_r$). The digital outputs of these RF chains are processed by a digital processor.}}
     \label{fig:SystemModel2}
\end{figure*}

\subsection{HRIS-Empowered Wireless Communications}
\label{subsec:CommSystem}
The most common application of conventional \acp{ris} is to facilitate communications between UTs and a \ac{bs} \cite{huang2019reconfigurable_all}, by shaping information-bearing signals over-the-air to overcome harsh non-line-of-sight conditions and improve coverage. In such setups, the \ac{ris} tunes its meta-atoms to generate favourable signal propagation profiles. To achieve this, the \ac{bs} maintains a control link with the digital controller of the \ac{ris}, where the latter sets the configuration of each meta-atom according to the control messages received by the \ac{bs}. This setup can be also extended to multiple \acp{ris} and cloud-controlled networks \cite{RISE6G_COMMAG_ALL,Alexandropoulos2022Pervasive}. 
This form of \ac{ris}-aided communications gives rise to several challenges. For instance, {the fact that the \ac{ris} must be remotely controlled by the \ac{bs} complicates its deployment and network management. Moreover,} in the absence of direct channels between the \ac{bs} and UTs, the former only observes the transmitted signals which propagated via the RIS-parameterized channel, namely, the composition of the \acp{ut}-\ac{ris} channel, the \ac{ris} reflection configuration, and the \ac{ris}-\ac{bs} channel. The fact that one cannot disentangle this combined channel implies that the \ac{bs} cannot estimate its individual components, but only the composite form \cite{wang2020channel_all}. This property does not only reflect on the ability to estimate the channels, but 
also limits the utilization of some network management tasks, such as wireless localization and sensing. 

We envision HRISs to be {utilized at least for the same purposes as} conventional reflective \acp{ris}. This includes, for instance, the typical application of coverage extension and connectivity boosting, by modifying the propagation profile of information-bearing \ac{em} waves, as illustrated in Fig.~\ref{fig:SystemModel2}. Similar to setups involving a conventional \ac{ris} \cite{huang2019reconfigurable_all}, the \ac{bs} {can maintain a dedicated control link with the HRIS, though the latter can also operate independently due to its sensing capability. For instance, the HRIS can use its collected estimations for the AoAs of intended impinging signals~\cite{locrxris_all} to decide its phase configuration, thus, relieving its dependence on external configuration and control~\cite{moustakas_2023_all}.
Nonetheless, when such a control link is present, it should not be used solely for unidirectional control messages from the \ac{bs} to the metasurface, as in reflective \acp{ris}, but now the HRIS can also use it to convey some sensed information to the \ac{bs}. The fact that the UTs' transmissions are also captured by the HRIS can notably improve channel estimation, as discussed in the sequel, which in turn facilitates coherent communications. Furthermore, the sensed signal can be used for localization, RF sensing, and RF mapping \cite{Keykhosravi2022infeasible_all,RIS_smart_cities}. 

As opposed to using reflective RISs that incorporate a few dedicated receive elements \cite{taha2021enabling_ALL} within their structure for channel estimation, HRISs profit from their entire surface to sense part of the impinging signal during periodic acquisition phases, while fully reflecting during data transmission. They can be also deployed for simultaneous channel estimation and data communication (similar to recent full-duplex-enabled systems~\cite{FD_SCDC,alexandropoulos2022fullduplex}) via blind estimation techniques. It is noted that the concept of receiving (semi-passive) RISs~\cite{alexandropoulos2020hardware,Tsinghua_RIS_Tutorial_ALL} can be implemented with a special case HRIS, having all its meta-atoms configured to fully sense.
An additional key difference between wireless networks empowered by HRISs and solely reflective \acp{ris} stems from the fact that the former do not reflect all the energy of the incident signal, when the HRIS is also configured to sense, since a portion of it is sensed and collected by the HRIS local processor. Nevertheless, this may degrade the \ac{snr} at the \ac{bs}. 

The potential benefits of HRISs over conventional ones come with additional energy consumption and cost. While the proposed meta-atom design in Fig.~\ref{fig:HybridAtom_v01} is passive, as that of solely reflective \acp{ris}, HRISs require additional power for locally processing the sensed signals. The utilization of RF chain(s) and the additional analog combining circuitry, which are not required by conventional \acp{ris}, is translated into increased cost. While explicitly quantifying this increased cost and power consumption is highly subject to the specific implementation~\cite{HRIS_intensity}, it is emphasized that an HRIS is still an \ac{ris}, and not an active multi-antenna relay. Namely, it does not involve a power-consuming wireless transmission mechanism for amplifying and transmitting its received signals, and is thus expected to maintain the cost and power gains of metasurfaces over traditional relays \cite{huang2019reconfigurable_all}. An additional challenge associated with HRISs stems from the optional need for a bidirectional control link between itself and the \ac{bs}. This link can be used for making HRIS's observations available to the BS (otherwise, they remain locally) to support higher throughput compared to the unidirectional control link utilized by conventional \acp{ris}.

\subsection{Model for Simultaneous Reflection and Sensing}
\label{subsec:Model}
A simple, yet generic, model for the HRIS operation can be obtained by considering its reflective and sensing capabilities individually. As discussed in Section~\ref{sec:Hybrid}, the HRIS tunable reflection profile is similar to conventional \acp{ris}, where it is typically modeled as controllable phase shifters. However, unlike conventional \acp{ris}, in HRISs, 
%
the reflectivity amplitude for each meta-atom is determined by a design knob that can be adjusted by changing the amount of coupling to the sensing circuitry. This design knob is modeled by a parameter $\rho_n\in[0,1]$, representing the portion of the signal energy being reflected by each $n$-th hybrid meta-atom. This parameter is related to the amount of coupling between the meta-atom and the sampling waveguide, which essentially determines how much loss the HRIS introduces compared to solely reflective RISs. Note that, if the coupling is too small, the sensed signal at the HRIS becomes susceptible to noise, while when it is too high, the signal reflected is considerably attenuated. As a result, it is useful to consider the reflectivity amplitude a design knob and optimize its acceptable range given a desired wireless system operation. 

The second part of the HRIS model describes the relationship between the impinging waveform at the HRIS and the field values picked up by its sensing circuitry. The exact details of this model are heavily dependent on the specific implementation (see Section \ref{sec:Future} for more details). For simplicity, one can consider this relationship to be as simple as the non-reflected part (i.e., $1-\rho_n$) multiplied by a phase component. Finally, as in \acp{dma} \cite{shlezinger2020dynamic_all}, 
the RF chain(s) via which the sensed signals are forwarded to the digital processor are connected to the waveguides, and thus, the sensed signal path can be viewed as a hybrid analog and digital receiver. 

The resulting HRIS operation model is illustrated in the right part of Fig.~\ref{fig:SystemModel2}, where one can configure the portion of the reflected energy for each meta-atom, as well as the phase shifting carried out at the reflected and received signal paths. We note that the HRIS may be configured to realize purely passive \acp{ris}~\cite{huang2019reconfigurable_all} (by setting $\rho_n =1$ for each $n$-th element) or the partially receiving RISs in~\cite{taha2021enabling_ALL} (by setting $\rho_n =0$ for some $n$ values and $\rho_{n'} =1$ for the remaining $n'$ values) or semi-passive \acp{ris}~\cite{alexandropoulos2020hardware,Tsinghua_RIS_Tutorial_ALL} (by setting $\rho_n =0$ for each $n$-th element). Accordingly, any performance achievable with the latter \acp{ris} is also achievable with an HRIS. However, HRISs provide increased flexibility, as the operation of their meta-atoms can be configured based on any current conditions, which greatly facilitates its self-automation and adaptivity~\cite{zhang2023channel_all}.
In practice, the parameters of the hybrid meta-atoms are typically coupled, implying that one cannot tune them arbitrarily. However, this relatively simple model allows to provide an understanding of what one can gain by properly tuning these HRIS parameters, as detailed in the following section.  

{
\begin{figure}
    \centering
    \includegraphics[width=1.06\columnwidth]{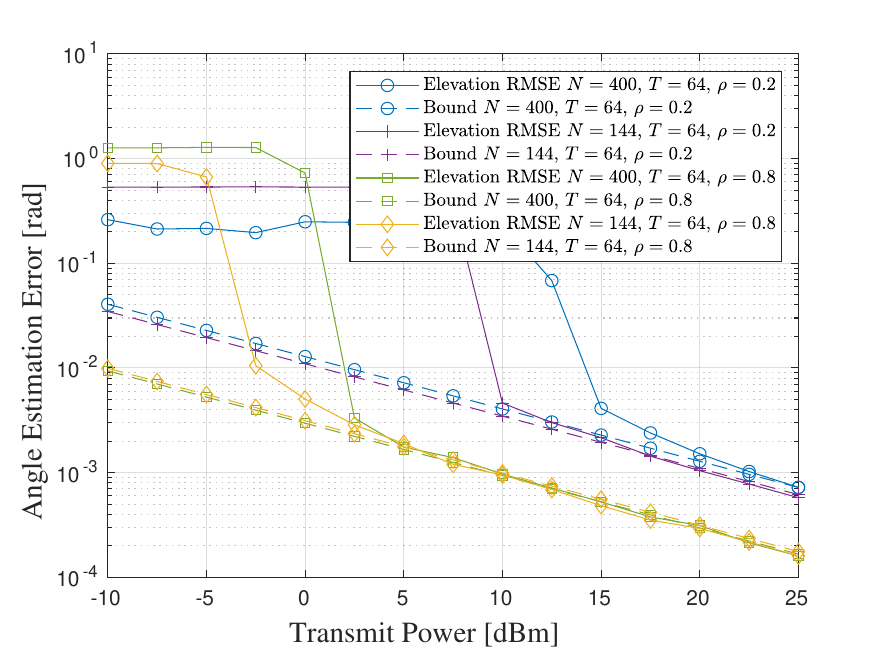}
    \vspace{-0.2cm}
    \caption{\small{Elevation AoA estimation performance of an HRIS with $N\in\{144,400\}$ meta-atoms configured with coupling coefficients $\rho_n=\rho \in \{0.2,0.8\}$ $\forall$$n=1,2,\ldots,N$ using $T=64$ sensed samples (measurements). The Cram\'{e}r Rao lower bounds of the estimations are also included.}}
    \label{fig:tradeoff}
\end{figure}
}

\subsection{Channel Estimation with HRISs}
\label{subsec:Estimation}
We next investigate the contribution of the sensing capability of HRISs in two indicative applications: \textit{i}) Estimation of the AoA of an impinging signal; and \textit{ii}) Estimation of the end-to-end channel matrices in multi-user uplink MIMO communications. The former is a key to enabling self-configuring HRISs~\cite{moustakas_2023_all}, alleviating their dependence on external control, and thus, notably facilitating deployment compared with conventional \acp{ris}. The latter highlights how, by sensing a portion of the impinging signal, an HRIS can estimate the composite \acp{ut}-HRIS channel and contribute to estimating the HRIS-\ac{bs} channel (see Fig.~\ref{fig:SystemModel2}). Such estimation is highly challenging when using a solely reflective \ac{ris}, since, in this case, only the channel outputs at the \ac{bs} can be observed~\cite{hu2019two_all}, and the common practice is to resort to estimating the combined channel effect, i.e., the cascaded channel \cite{wang2020channel_all}, using pilot RIS reflection patterns.




\smallskip
{\bf AoA Estimation}: 
We consider the estimation of the AoA of an impinging signal in quasi free-space. For simplicity, we use a two-dimensional square metasurface with $N$ hybrid meta-atoms (as in Section~\ref{sec:Hybrid}) operating at  $19$ GHz (corresponding to wavelength $\lambda=15.70$ mm) with element spacing of $4$ mm ($\approx\lambda/4$) apart. The design frequency is similar to that used in \cite{sleasman2016microwave_all} and is considered for demonstration purposes. All meta-atoms are coupled to a single waveguide with the common coupling coefficient $\rho$, which is connected to a single reception RF chain. Directive analog combining is used to feed this chain's input, while its digital outputs after $T$ received samples are used to estimate the AoA using a maximum likelihood estimator.  
While the AoA is comprised of azimuth and elevation angle, we focus here on the elevation AoA.

The numerically evaluated AoA performance in root \ac{mse}, compared with the Cram\'{e}r-Rao estimation lower bound, is illustrated in Fig.~\ref{fig:tradeoff} for $N\in \{12^2, 20^2\}$, $\rho_n=\rho\in \{0.2, 0.8\}$ $\forall$$n=1,2,\ldots,N$, and $T=64$. As depicted, the AoA of the impinging signal can be accurately estimated even when the portion of the signal being sensed is merely $20\%$ of that incoming. However, as expected, by increasing the meta-atoms' coupling to the waveguide, i.e., by sensing a more dominant portion of the impinging signal, the estimation performance improves. It is also demonstrated that a smaller HRIS yields better performance, which is attributed to the fact that wider beams facilitate AoA estimation. These results demonstrate the ability of HRISs to reliably identify the AoA of an impinging signal, even when only a small portion of it is being sensed, while the remainder is reflected. 

\color{black}

\smallskip
{\bf Full Channel Estimation}: 
To demonstrate the HRIS channel estimation capability, we consider the uplink system of Fig.~\ref{fig:SystemModel2} with $K$ single-antenna \acp{ut} and an H\ac{ris} with $N$ hybrid meta-atoms. 
The metasurface has $N_r$ reception RF chains and maintains an out-of-band unidirectional control link with the \ac{bs}. A simple strategy to estimate the individual channels is to assign the H\ac{ris} to estimate the \acp{ut}-\ac{ris} channel via its sensed observations, and to forward this estimate to the \ac{bs} over the control link (wired or wireless)~\cite{RISE6G_COMMAG_ALL}, while changing the phase configuration between pilot symbols as in \cite{wang2020channel_all}. This strategy has similar latency with typical channel estimation schemes for conventional RISs~\cite{Tsinghua_RIS_Tutorial_ALL}. The fact that the H\ac{ris} also reflects while sensing, allows the \ac{bs} to estimate the H\ac{ris}-\ac{bs} channel from its observed reflections, effectively re-using the transmitted pilots for estimating both hops' channels. According to the latter strategy, the H\ac{ris} estimates the $NK$ coefficients comprising the \acp{ut}-\ac{ris} channel; the measurements are obtained via the $N_r$ digital signal paths. In the absence of noise, the H\ac{ris} is able to recover the \acp{ut}-H\ac{ris} channel from $NK / N_r$ pilots, while the \ac{bs} requires at least $N$ pilots to recover the H\ac{ris}-\ac{bs} channel fuzzing its observations and the estimate provided by the HRIS via the control link. For instance, when $K=8$ \acp{ut} communicate via an H\ac{ris} comprising $N=64$ meta-atoms, all attached to $N_r=8$ RF chains, both individual channels can be recovered from merely $64$ pilots. State-of-the-art methods for conventional \acp{ris} would require over $90$ pilot transmissions for recovering solely the cascaded channel for a fully digital $16$-antenna \ac{bs} \cite{wang2020channel_all}. This simplistic strategy for channel estimation is beneficial, not only in noise-free channels, but also in noisy setups. However, one needs to also account for the fact that, the power splitting between the reflected and sensed waveforms, carried out by the H\ac{ris}, affects the resulting \acp{snr} at both the metasurface and \ac{bs}. In particular, in noisy setups where the H\ac{ris} estimates the composite \acp{ut}-H\ac{ris} channel locally and forwards it to the \ac{bs}, there is an inherent trade-off between the accuracy in estimating each of the individual channels. This is dictated by how the H\ac{ris} is configured to split the power of the incident wave between its reflected and absorbed portions.
\begin{figure}
    \centering
    \includegraphics[width=1.08\columnwidth]{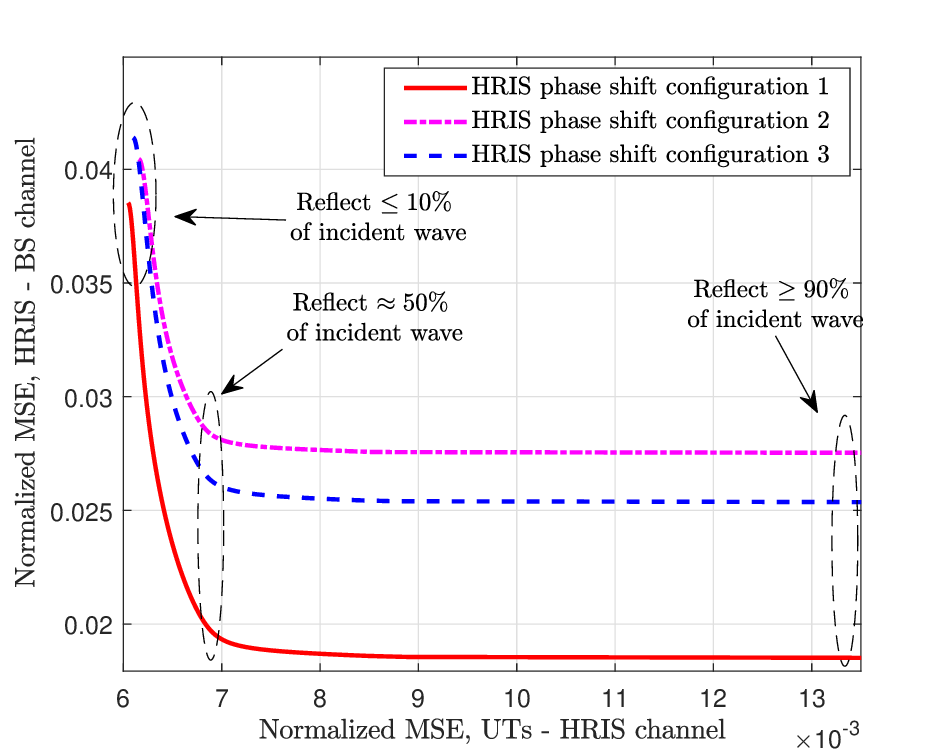}
    \vspace{-0.2cm}
    \caption{\small{Normalized \ac{mse} performance in recovering the composite \acp{ut}-HRIS channel at the HRIS and the HRIS-\ac{bs} channel at the \ac{bs}, considering $30$ dB transmit SNR as well as different power splitting values $\rho$ ($\rho_n=\rho$ $\forall$$n=1,2,\dots,64$) and phase configurations. The fully digital BS is equipped with $16$ antenna elements, there exist $8$ \acp{ut} in the system, and the HRIS possesses $8$ reception RF chains.}}
    \label{fig:tradeoff}
\end{figure}

The aforedescribed trade-off between the \ac{mse} performance when estimating the \acp{ut}-H\ac{ris} channel at the H\ac{ris} and the H\ac{ris}-\ac{bs} channel at the \ac{bs} for different values of the common power splitting parameter $\rho$ is depicted in Fig.~\ref{fig:tradeoff}. 
Each curve corresponds to a different random setting of the individual phase shift at each meta-atom. The simulation setup consists of a \ac{bs} with $16$ antennas serving $8$ \acp{ut}, assisted by an H\ac{ris} with $64$ elements and $8$ RF chains. The \acp{ut} are uniformly distributed in a cell of $10$ meter radius, where the H\ac{ris} is located at the top edge of the cell, and at distance $50$ meters from the \ac{bs}. The \acp{ut} transmit $70$ pilot symbols for channel estimation, which are received at the \ac{bs} after being reflected by the H\ac{ris} with \ac{snr} $30$ dB. It is observed, in Fig. \ref{fig:tradeoff}, that there is a clear trade-off between the accuracy in estimating each of the individual channels, which is dictated by how the H\ac{ris} splits the power of the impinging signal. While the \ac{mse} values depend on the HRIS phase configuration, we observe that increasing the portion of the signal that is reflected in the range of up to $50\%$ notably improves the ability to estimate the H\ac{ris}-\ac{bs} channel, while having only a minor effect on the \ac{mse} of the \acp{ut}-H\ac{ris} channel estimation. However, further increasing the amount of power reflected, notably degrades the \ac{mse} in estimating the \acp{ut}-H\ac{ris} channel, while hardly improving the accuracy of the H\ac{ris}-\ac{bs} channel estimation.

The ability to sense the individual channels at the HRIS side facilitates channel estimation, especially in wideband multi-user multi-antenna systems, where cascaded channel estimation requires prohibitive overhead~\cite{wang2020channel_all}. In addition, it enables RF sensing, localization, and RF mapping, whose role is envisioned to be prominent in 6G~\cite{Samsung}. 
Interestingly, it can be beneficial even when one is interested in the cascaded channel. To show this, in Fig. \ref{fig:comparison}, we demonstrate the \ac{mse} in estimating the cascaded channel for the setup in Fig.~\ref{fig:tradeoff} with an H\ac{ris}, in comparison to the method in \cite{wang2020channel_all} for solely reflective \acp{ris}. Two different HRIS configurations were considered: a fixed setting, where the coupling parameters $\rho_n$'s were tuned to reflect on average {$50\%$} of the incoming signal, and a setting where these parameters were optimized using first-order optimization of the MSE objective employing automatic differentiation tools~\cite{zhang2023channel_all}. 
As observed, the sensing capability of H\acp{ris} translates into improved  cascaded channel estimation, which is more prominent when $\rho_n$'s are optimized, depending on how many RF chains the HRIS possesses. This behavior stems from the fact that, the channel estimation procedure adopted here is geared towards estimating the individual channels without imposing any structure on them, and enables the HRIS to estimate the \acp{ut}-H\ac{ris} locally. While, when the number of RF chains is small, the error induced in such local estimation reflects on the cascaded channel recovery, when employing $5$ RF chains or more, notable performance improvements are observed, though at the cost of higher power consumption and hardware complexity. 
\begin{figure}
    \centering
    \includegraphics[width=1.08\columnwidth]{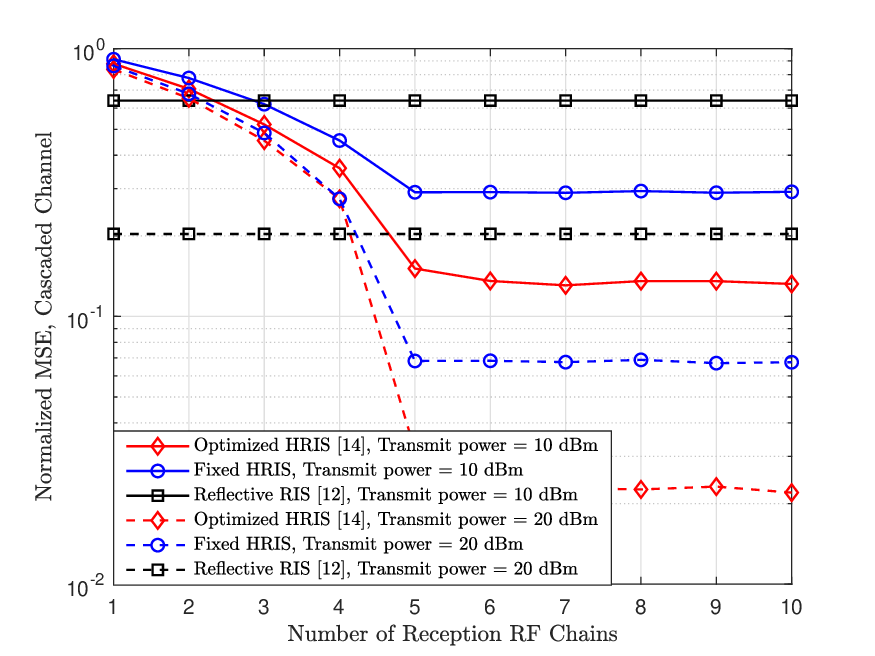}
    \vspace{-0.2cm}
    \caption{\small{Normalized \ac{mse} performance of the cascaded channel estimation as a function of the number of reception RF chains at the HRIS, considering fixed and optimized $\rho_n$'s (via the approach in~\cite{zhang2023channel_all}) for two different transmit power values in dBm. The respective estimation performance using a solely reflective RIS, via the scheme of \cite{wang2020channel_all}, is also included.}}
    \label{fig:comparison}
\end{figure}

\section{Research Challenges and Opportunities}
\label{sec:Future}
The design of H\acp{ris} (Section~\ref{sec:Hybrid}) and their representative applicative capabilities (Section~\ref{sec:Comm}) provide an indication of the concept's feasibility and gains. Their ability to carry out reflection and sensing simultaneously gives rise to a multitude of research challenges and opportunities. Some important investigations, which we expect to play a key role in unveiling the potential of such metasurface architectures in future smart wireless environments, are next discussed. 

{\bf Fundamental Limits:} The inclusion of RF sensing capabilities at metasurfaces considerably changes the operation of metasurface-assisted wireless communication systems. This motivates the characterization of the fundamental limits of H\ac{ris}-empowered sensing-aided communications in terms of achievable rate. Such an analysis can quantify the added value of H\acp{ris} compared to solely reflective RISs, in a manner that is invariant of the operation of the network end entities. 

{\bf Operation Protocols:} HRISs provide additional network design degrees of freedom compared to conventional RISs and inherently facilitate integrated sensing and communications schemes, motivating novel algorithmic designs for their exploitation. Their RF sensing, and the relevant computing capabilities, enable signal processing at the metasurface side (e.g., direction estimation, localization, and RF mapping), which can be used locally to enable their self-configuration, thus significantly reducing the network overhead for HRIS optimization, or shared to the network for management purposes. In the latter case, efficient low-latency control protocols are needed (possibly among multiple HRISs and BSs), while considering rate-limited in- or out-of-band control links, as well as self-configuring HRISs.  

{\bf HRIS Modeling:} The performance evaluation 
detailed in Section~\ref{sec:Comm} is based on the presented simplified model for the HRIS dual operation. In practice, the EM response of each hybrid meta-atom is expected to exhibit a more complex configurable profile, including coupling between its parameters (i.e., coupling and phase shifting coefficients) as well as between different elements. Such a physics-driven characterization will allow to more faithfully evaluate H\acp{ris}. An additional critical aspect which should be characterized is the HRIS excessive power consumption and cost compared to reflective \acp{ris}. While such an analysis is expected to be highly implementation dependent, it will help understanding the price associated with the H\ac{ris} gains.

{\bf Hardware Designs:} The presented proof-of-concept, which is based on \cite{alamzadeh2021reconfigurable_all}, is an important first step in demonstrating  HRIS hardware. Nonetheless, realizing such metasurfaces for wireless operations still requires a large body of experimental efforts and hardware designs, from low up to THz frequencies~\cite{RIS_THz}. Akin to purely reflective \acp{ris}, the H\ac{ris} design requires investigations into the number and pattern of meta-atoms to be implemented as hybrid and their tuning mechanism, as well as the overall size of the metasurface and its phase configuration capabilities (e.g., beam steering span, grating lobes, and sidelobes). All other studies related to forming reflection patterns (as done for conventional \acp{ris}) are also relevant to HRISs. 

Moreover, using the implementation in Fig.~\ref{fig:HybridAtom_v01}, the complex amplitude sensed at the waveguides does not necessarily exhibit an one-to-one relationship with that at each meta-atom. This happens because the wave coupled by each of them into the substrate layer travels along it, gets reflected from boundaries, and can couple to all other sampling waveguides. This overall effect can be considered as information multiplexing, and can be pre-characterized into a sensing matrix, which may be used for channel estimation. In an alternative implementation to Fig.~\ref{fig:HybridAtom_v01}, each hybrid meta-atom can be directly connected to a sampling waveguide; the wave incident on different meta-atoms will now have minimal coupling. In this case, one has a nearly one-to-one relationship between the wave impinging at the meta-atom and the one at the sampling waveguide, and the resulting multiplexing factor can be used for noise mitigation.



\section{Conclusion}
\label{sec:Conclusions}
In this article, we reviewed the emerging concept of HRISs for smart wireless environments applications. In contrast to solely reflective RISs, HRISs can simultaneously reflect a portion of the impinging signal in a controllable manner, while sensing the other portion of it. We presented an HRIS design and discussed a full-wave EM proof of concept, showcasing the concept's hardware feasibility. We highlighted the possible operations of H\ac{ris}-empowered wireless communication systems, provided a simplified model for the HRIS reconfigurability, and evaluated their capability to locally identify AoAs and to facilitate channel estimation, in comparison with solely reflective RISs. We discussed several research challenges and opportunities arising from the HRIS concept, which are expected to pave the way in unveiling the potential of this technology for 6G wireless systems.

\bibliographystyle{IEEEtran}
\bibliography{IEEEabrv, references}

\end{document}